\documentclass{iopart}

\usepackage{graphicx}
\usepackage{amsfonts}

\begin{document}

\newcommand{\myL}{\mathcal{L}}
\newcommand{\del}{\nabla}
\newcommand{\Real}{\mathbb{R}}
\newcommand{\interp}{I_{2h}^h\,}
\newcommand{\prolong}{I_{2h}^h\,}
\newcommand{\restrict}{I_{h}^{2h}\,}
\newcommand{\mkeq}[1]{\begin{equation}#1\end{equation}}

\title[High order multigrid methods]
    {High order convergent multigrid methods on domains containing holes
for black hole initial data}

\author{Vishnu Natchu and Richard A. Matzner}

\address{Center for Relativity,
        University of Texas at Austin,
        Austin, Texas 78712 USA}
\ead{vishnu@physics.utexas.edu, matzner2@physics.utexas.edu}

\begin{abstract}
It is well known that multigrid methods are optimally efficient for
solution of elliptic equations ($O(N)$), which means that effort is
proportional to the number of points at which the solution is
evaluated).  Thus this is an ideal method to solve the initial
data/constraint equations in General Relativity for (for instance)
black hole interactions, or for other strong-field gravitational
configurations.  Recent efforts have produced finite difference
multigrid solvers for domains with holes (excised regions).  We
present here the extension of these concepts to higher order (fourth-,
sixth- and eigth-order).  The high order convergence allows rapid
solution on relatively small computational grids.  Also, general
relativity evolution codes are moving to typically fourth-order; data
have to be computed at least as accurately as this same order for
straightfoward demonstration of the proper order of convergence in the
evolution.

Our vertex-centered multigrid code demonstrates globally
high-order-accurate solutions of elliptic equations over domains
containing holes, in two spatial dimensions with fixed (Dirichlet)
outer boundary conditions, and in three spatial dimensions with {\it
  Robin} outer boundary conditions.  We demonstrate a ``real world''
3-dimensional problem which is the solution of the conformally flat
Hamiltonian constraint of General Relativity.  The success of this
method depends on: a) the choice of the discretization near the holes;
b) the definition of the location of the inner boundary, which allows
resolution of the hole even on the coarsest grids; and on maintaining
the same order of convergence at the boundaries as in the interior of
the computational domain.
\end{abstract}


\section{Introduction}
Solving Einstein's equation in 3+1 form \cite{ADM,York} requires
that a set of elliptic (or quasi-elliptic) constraint equations be
satisfied at all times.  The Bianchi identities ensure that, given
a set of initial data which analytically satisfy the constraints,
the subsequent analytically evolved variables will also satisfy the
constraints.   In numerical solutions of Einstein's equations,
however, the constraints are not preserved exactly.  Thus errors
will arise as the simulation proceeds, and the extent to which the
numerical solutions actually reflect the true solutions of the
analytic equations is an ongoing area of research (e.g.,
\cite{Siebel,Calabrese,Tiglio,Yoneda,Apples}).  Apart from the issue
of the accuracy of the solutions obtained, there is also the problem
of numerical (in)stability, may in some cases be related to the lack
of preservation of the constraints.  There is active
interest in constrained general relativity evolution schemes (which
repeatedly solves the constraint equations during the
evolution\cite{Erik,DaveMeier,AndersonMatzner}).

The multigrid method is a particularly attractive method to solve
elliptic equations because it is an optimal method, i.e. it requires
only $O(N)$ operations, where $N$ is the number of unknowns
(proportional to the number of locations at which knowledge of the
independent variable is required).  Furthermore, the multigrid method
is not especially difficult to implement.  We study the multigrid
problem on domains with holes, because while many formulations now can
produce moving black holes without excising the holes, evolution with
excision in a necessary requirement for fully studying and
understanding black hole evolution.  For instance, posing data for black
holes with Kerr parameter $a$ very near unity are straightforward in
excised data, but very difficult to achieve in non-excised data (such
as {\it puncture} data\cite{dain}).

It has long been thought infeasible
in practice to obtain generic results which have everywhere the
same order of accuracy as the finite difference scheme employed,
for domains containing holes. This argument was
refuted in \cite{HawleyMatzner}
for 3D solutions of second-order accuracy. This paper addresses the problem
for domains with coordinate-spherical holes in
the context of higher order accuracy. Multigrid provides
a simple, robust method to achieve solutions in such cases.

If the multigrid solver is to be used to provide initial data for
a numerical evolution code, the error of the initial solution need
only be below the truncation error of the finite difference scheme
used for evolution \cite{Mattcomment}.  However high order methods
are still extremely valuable in this context because they require much
smaller computational resources for the same accuracy. Thus this
high order approach may dramatically lower the elliptic - solution
load in constrained evolutions.
Furthermore, to correctly demonstrate the convergence of evolution codes
(now typically written in terms of fourth-order stencils), one must have
data of
at least the same order of convergence.

Finite difference techniques to solve the constraint
equations are a standard activity in relativity, and the
properties of the constraint equations are well known (for the
Euclidean, constant-mean-curvature background and for the signs of
the solutions we use here \cite{Choquet,Dain2001,Maxwell}).
We focus on the benefits of high-order multigrid
implementation.
The validity of our results will be established by
demonstrating controlled high order convergence to known exact solutions.

Section \ref{overview} assumes familiarity with the multigrid method,
but a small overview is presented.
Section \ref{2dsec} describes our new scheme for inner boundary points
and presents the results for a simple problem in a 2D
domain. Section \ref{3dsec} extends this to 3D.
Section \ref{conc} summarizes our results.

\section{Overview of Multigrid\label{overview}}
Application of the multigrid method \cite{ABrandt},
on domains with holes --- or on domains
with ``irregular boundaries'' in general --- has received only
modest attention
\cite{Johansen}, \cite{Udaykumar}; these are cell centered finite
difference codes.  The ``BAM'' code
\cite{BAM} provides access to the multigrid method in
numerical relativity. It is a vertex centered second-order code and
features the ability to handle domains
with multiple holes, though with restrictions on hole placement. Hawley and
Matzner \cite{HawleyMatzner} introduced a second-order vertex-centered
multigrid code that
accommodates arbitrary placement of excised holes.
The innovation in this current paper is solution to higher convergence order
(we
give fourth- and sixth-
order examples. and some preliminary, eighth-order results) which
dramatically reduces the time and storage requirements of the elliptic
solution.

The multigrid scheme \cite{ABrandt}
has received considerable attention in the literature, and is the
subject of numerous articles, conferences, reviews and books (e.g.,
\cite{HackTrot,Hackbusch,Stueben,Wesseling,WesselingBook,Briggs}). It is
essentially a clever means of
eliminating successive wavelength-components of the error via the
use of relaxation at multiple spatial scales.

Here we give a very brief overview of the multigrid method, following
the notes by Choptuik \cite{Mattsnotes}.  (Introductions to
multigrid
applications in numerical relativity are also found in Choptuik and Unruh
\cite{ChopUnruh} and Brandt \cite{BrandtSchw}.) We want to solve
a continuum differential equation $\myL u = f$, where $\myL$ is a
differential operator, $f$ is some right hand side, and $u$ is the
solution we wish to obtain.  We discretize this differential
equation into a {\it difference} equation on some grid (or lattice) with
uniform spacing $h$:
\begin{equation}
  \myL^h u^h = f^h,
  \label{differenceeq}
\end{equation}
where $u^h$ is the {\em exact} solution of this discrete equation,
and $\lim_{h\rightarrow 0} u^h = u$.
The discretization $h$ refers to the finest grid, of
a hierarchy of vertex-centered grids.
Each grid at multigrid level $l$ is a square lattice having $2^l +
1$ grid points along each edge.  The grids have uniform spacing
$h_l = 2^{-l}$ in both $x$ and $y$ directions, and the grid points
are denoted with indices $i$ and $j$ in the $x$ and $y$ directions,
respectively, e.g.,  $u(ih_l,jh_l) \simeq \tilde{u}^h_{i,j}$.

Rather than attempting to solve Eq. (\ref{differenceeq}) directly via a
costly matrix inversion, we apply an iterative solution
method.  At any step in the iteration, we have an
approximate solution
$\tilde{u}^h \simeq u^h$.
In this iterative algorithm, the original guess
$\tilde{u}^h_{\rm old}$
is brought closer to $u^h$ by applying some (approximate)
correction:
\mkeq{ \tilde{u}^h_{\rm new} := \tilde{u}^h_{\rm old} + \tilde{v}^h
  \label{uneweq}}

For nonlinear
operators the correction is via the Full Approximation Storage (FAS) method
\cite{HawleyMatzner,ABrandt,HackTrot,Hackbusch,Stueben,Wesseling,WesselingBook,Briggs}.

\subsection{V-Cycles and the Full Multigrid Algorithm}
The solution algorithm takes the form of
a {\em V-cycle}, in which we start with an initial
guess on the fine grid, at multigrid level $l_{\rm max}$
(the finest grid has $2^{l_{\rm max}}+1$ vertices along each edge).  Then we
perform some number of {\t smoothing sweeps}. A smoothing sweep is one
iteration of a relaxation solver for the elliptic equation. The effect
of such a step is to reduce the short wavelength error on the grid,
i.e. it ``smooths" the approximate solution.
At this point we wish to update Eq.(\ref{uneweq}).

To accomplish this the multigrid method introduces
$\restrict$, the {\it restriction} operator, which
maps values from the fine grid to the next coarser grid via some
weighted averaging operation, and  $\interp$, an interpolation or {\it
prolongation} operator,
which maps values from a coarse grid to the next finer grid via some
interpolation operation.
In the Full Approximation Storage multigrid method we express the correction
as

\mkeq{ \tilde{u}^h_{\rm new} := \tilde{u}^h_{\rm old} +
  \interp (u^{2h} - \restrict \tilde{u}^{h}).\label{FASupdate}}
Eq(\ref{FASupdate}) defines a {\it coarse grid correction} (CGC). 

In Eq(\ref{FASupdate}),  $u^{2h}$ is the exact solution to (\ref{differenceeq}) on the next
coarser grid.
However, we may approximately solve for the coarser grid
$\tilde{u}^{2h} \simeq u^{2h}$
by repeating Eq. (\ref{FASupdate}) on the coarse grid,
which then refers to the next (even) coarser grid.
We continue smoothing and restricting to coarser
grids until we arrive at a grid coarse enough to solve the resulting coarse
grid equation `exactly' (i.e., to machine precision),
at minimal computational
cost.
Equation (\ref{FASupdate}) can then be used to correct the next finer grid
solution.
We can thus  iteratively correct the solutions on each next finer grid,
perhaps with additional smoothing operations
performed before moving to each finer grid. One may carry out a number
of smoothing sweeps at a given refinement level {\it before} proceeding
to solve on the next coarser grid ({\it pre-sweeps}), and a
(perhaps different) number of {\it post-sweeps}
{\it after} solving on the next
coarser grid. We use the same numbers of pre- and the same number of
post- sweeps at every level, though the numbers could in principle be
varied at different levels.

On all grids except the coarsest grid, we only
smooth the error, and we solve the difference equation exactly
only on the coarsest grid. In practice we solve the coarse grid
difference equation by relaxation, which is cheap on the coarsest
grid. The entire process, as described is
called a V-cycle. A full solution may consist of one or more V-cycles.

\section{Solution of a Nonlinear Poisson Equation in 2D\label{2dsec}}

We solve the equation
\mkeq { {\partial^2 \over \partial x^2} u(x,y) +
       {\partial^2 \over \partial y^2} u(x,y)
       + \sigma u^2(x,y) = f(x,y),
\label{2dPoisson}
}
on a domain $\Omega$ with coordinate
ranges $[0,0]$ to $[1,1]$, and
subject to Dirichlet conditions at the outer boundary:
$u(x,y)|_{\partial \Omega_O} = 0$.
The function $f(x,y)$ is chosen such that the solution is
\mkeq{ u(x,y) = \sin(\pi x) \sin(\pi y),
\label{2dSine}}

\hfil\break
\noindent{\em Inner Boundary Conditions}\hfil\break

We have added features to handle holes in the domain.
In  \cite{HawleyMatzner} the inner boundary is given by the outer edge of
the points comprising
the excision mask.
This has the consequence that the
``size'' of the excision region (i.e., the area of the convex hull
of the data points comprising the mask) on finer grids is always
equal to or greater than the size of the excision region on coarser
grids.
Here we present a novel modification of the inner boundary algorithm:
at the inner boundary we insert a new set of points into the
grid, and apply Dirichlet boundary conditions (the exact solution) to these
points.
The boundary conditions are applied at each level. This has the effect
of keeping the size of the excised region essentially constant at all
levels.

\begin{figure}
\centering
\includegraphics[width=4in]{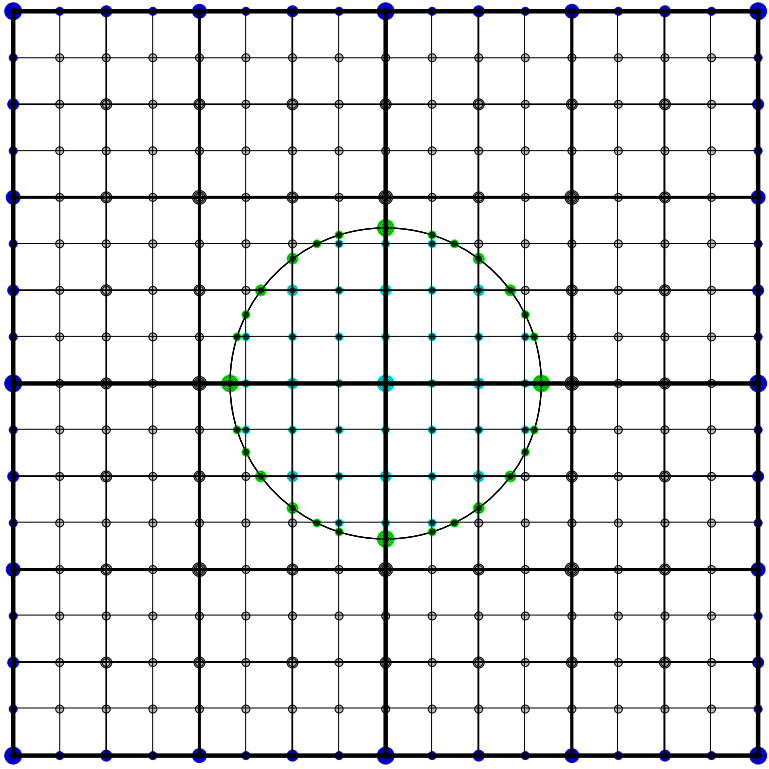}
\caption{
\label{fig_boundary}
Extra grid points points added at intersection of the inner boundary
and the grid lines.  Dirichlet boundary conditions are used to fill in
value for these points.
}
\label{fig_define_ex_reg}
\end{figure}

These extra points are added at each level at every location at which a
grid line intersects the exact inner boundary location.  The new
boundary
points specify the exact solution to the inner boundary conditions.  The
neighboring grid points use this new point in their smoothing
and stencil operations taking care to account for the irregular grid
size caused by adding this point.
Figure \ref{fig_boundary} illustrates this. Note that this has the important
result
that the hole is resolved on even the coarsest grid. We will find that this
approach dramatically reduces the solution error in domains with holes.
Although Figure \ref{fig_boundary} shows a circular excision centered 
on a grid point, we have 
verified that convergence is maintained even for arbitrary placement 
of the excision circle (or sphere, in 3D).

\hfil\break
\noindent{\em Smoothing Operations}\hfil\break

Except on the coarsest grid, we do not attempt to solve the difference
equation Instead we use a few sweeps of a relaxation method. This is called
{\it smoothing}.
A ``red-black'' Gauss-Seidel Newton iteration is used to
relax the solution(smooth the error). For all points that have a
regular spacing $h$ with their neighbors (interior points) this is the 
traditional method.  The values of $h$ differ by a
factor of
$2$ between ``adjacent" grid levels.  For these 2D examples we use a
second-order discretization.
On interior points, Eq(\ref{2dPoisson}) is discretized as

\begin{eqnarray}
        r_{GS} & = & h^{-2} \left(
                           \tilde{u}_{i+1,j} + \tilde{u}_{i-1,j}
                         + \tilde{u}_{i,j+1} + \tilde{u}_{i,j-1}
                         - 4 \tilde{u}_{i,j} \right) \cr
          &\mbox{}& + \sigma  \tilde{u}^2_{i,j} -
                         f_{i,j}
\end{eqnarray}

\begin{equation}
          \tilde{u}^{\rm new}_{i,j} = \tilde{u}^{\rm old}_{i,j}
                            - { r_{GS} \over
                            2 \sigma \tilde{u}_{i,j} - 4h^{-2}}.
\label{2dgaussSeidel}
\end{equation}
But for points where the spacings are not regular, we introduce new notation
here. Define $h_{i+{1\over 2}, j}$ as the spacing in the $x$ direction
between points at $i,j$ and $i+1,j$, similarly $h_{i-{1\over 2}, j}$ is the
spacing
between  $i,j$ and $i-1,j$.  Note that there can be multiple substitutions
for an
excised grid point, so we choose a particular one, and the associated $h_{i
\pm {1\over 2}, j}$.

\begin{eqnarray}
 r_{GS} & = & \frac{2}{h_{i+\frac{1}{2},j}+h_{i-\frac{1}{2},j}}
 \left(\frac{(\tilde{u}_{i+1,j} -
   \tilde{u}_{i,j})}{h_{i+\frac{1}{2},j}} + \frac{(\tilde{u}_{i-1,j}
   - \tilde{u}_{i,j})}{h_{i-\frac{1}{2},j}}\right) \cr
&\mbox{} & + \frac{2}{h_{i,j+\frac{1}{2}}+h_{i, j-\frac{1}{2}}}
\left(\frac{(\tilde{u}_{i,j+1} -
   \tilde{u}_{i,j})}{h_{i, j+\frac{1}{2}}} + \frac{(\tilde{u}_{i,j-1}
   - \tilde{u}_{i,j})}{h_{i, j-\frac{1}{2}}}\right) \cr
 &\mbox{}& + \sigma  \tilde{u}^2_{i,j} -
                         f_{i,j}\label{relax1}\\
  f_u & = & (\frac{1}{h_{i+\frac{1}{2}, j}} +
  \frac{1}{h_{i-\frac{1}{2}, j}})\frac{2}
  {h_{i+\frac{1}{2}, j}
    + h_{i-\frac{1}{2}, j}} \cr
  &\mbox{}& +
  (\frac{1}{h_{i, j+\frac{1}{2}}} +
  \frac{1}{h_{i, j-\frac{1}{2}}})
  \frac{2}{h_{i, j+\frac{1}{2}}
    + h_{i, j-\frac{1}{2}}}\label{relax2}\\
  \tilde{u}^{\rm new}_{i,j} & = & \tilde{u}^{\rm old}_{i,j}
                            - { r_{GS} \over
                            2 \sigma \tilde{u}_{i,j} - f_u}\label{relax3}
\end{eqnarray}
Note that because of the unequal spacing, Eqs(\ref{relax1} - \ref{relax3})are in fact
only {\it first-order}
near the inner excision boundary. For 2D we find that this does not
contaminate the high order convergence of the solutions.

\hfil\break
\noindent{\em Restriction and Prolongation Operators}\hfil\break

The restriction operator $\restrict$ we use is the so-called
``half-weighted'' average on normal interior points, in which coarse grid
values (indexed by $I$ and $J$ for clarity) are a weighted average
of the fine grid values over a nearby region of the physical domain:
\begin{equation}
           \tilde{u}^{2h}_{I,J} = \restrict \tilde{u}^h =
              {1\over 2} \tilde{u}^h_{i,j} +
              {1\over 8} \left[
                   \tilde{u}^h_{i+1,j} + \tilde{u}^h_{i-1,j} +
                   \tilde{u}^h_{i,j+1} + \tilde{u}^h_{i,j-1} \right],
\label{2drestrict}
\end{equation}
where $i = 2I - 1$ and $j = 2J -1$.

For the prolongation operator $\prolong$, we use simple bilinear
interpolation.

\subsection{2D Results}\hfil\break

\begin{figure}
\centering
\centerline{\includegraphics[width=3.0in,angle=90,height=3.5in]{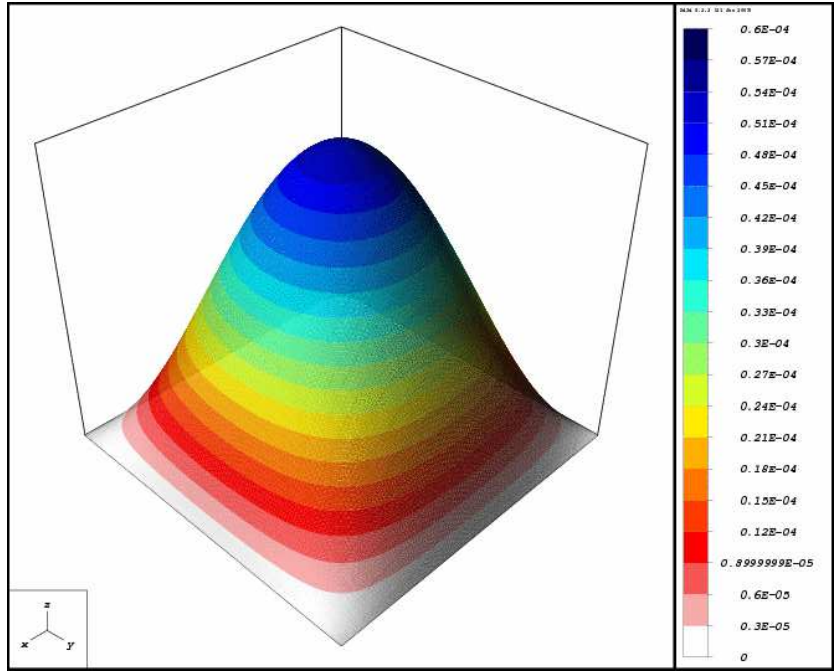}}
\caption{
A plot of the solution error $e = |u - \tilde{u}^h|$ for the 2D case.  We
present a second-order
solution with no holes to provide a metric for comparison for the
rest of the results.
$l_{\rm max} = 7$
(the finest grid has $2^7+1=129$ vertices along an edge), and
$l_{\rm min} = 2$
(the coarsest grid has $2^2+1=5$ vertices along an edge).
2 V-cycle and 2 pre- and 2 post-CGC smoothing sweeps.
}
\label{fig_2dplot_nohole}
\end{figure}

\begin{figure}
\centering
\centerline{\includegraphics[width=3.0in,angle=90,height=3.5in]{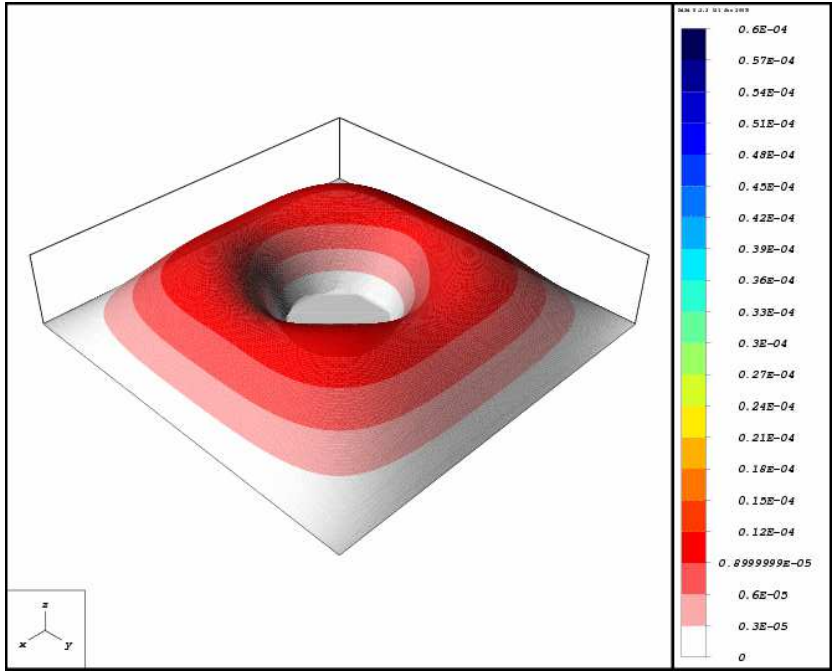}
}
\caption{
Solution error $e = |u - \tilde{u}^h|$.  A 2D simulation with a central
circular hole of radius $r_{\rm mask} = 0.129$ (second-order
discretization). Solution parameters: $l_{\rm max}=7$, $l_{\rm
min}=2$, 2 V-cycle, with 2 pre- and 2 post-CGC smoothing sweeps.
}
\label{fig_2dplot_hole}
\end{figure}

\begin{figure}
\centering
\centerline{\includegraphics[width=3.5in,angle=90,height=3.5in]{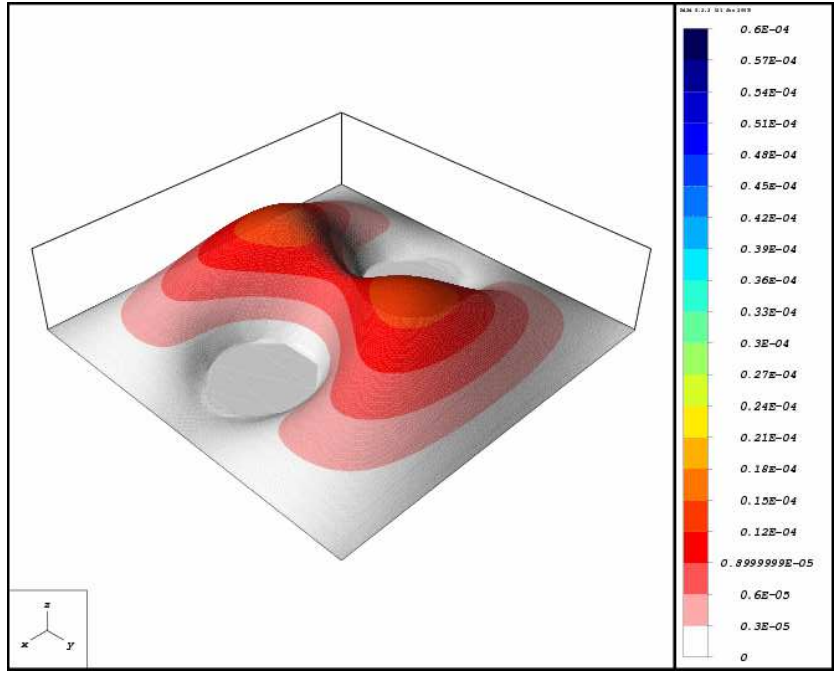}}
\caption{ Solution error for 2D simulation with two holes (second-order
discretization).
$l_{\rm max}=7$, $l_{\rm
min}=2$, 2 V-cycle and 2 pre- and 2 post-CGC smoothing sweeps.
\label{fig_2dplot_twohole}}
\end{figure}

We solve Eq(\ref{2dPoisson}) with $\sigma = 1$. We take $l_{max} = 7$
(the finest grid has $2^7+1=129$ vertices along an edge), and
$l_{min} = 2$
(the coarsest grid has $2^2+1=5$ vertices along an edge).
Figure \ref{fig_2dplot_nohole} is a plot of the error over the entire
domain, with no holes in the domain. Most implementations of multigrid with
irregular inner
boundaries end up with higher errors than ones without the holes, but
the extra inner boundary condition might also help ``tie the solution
down''.  Figure \ref{fig_2dplot_hole}
shows solution error for a domain with an excised region (and inner
boundary)
of radius $r=0.129$.  The scales of Figures \ref{fig_2dplot_nohole}
through \ref{fig_2dplot_twohole} are matched, so it is
easy to see that the error decreases. Also note that the error is very
tightly bounded close to the inner boundary; evidently the inner boundary
condition is reducing the maximum error in the domain. Figure
\ref{fig_2dplot_twohole} which shows the solution error when the domain
contains two holes, confirms this.

\begin{figure}
\centering
\centerline{
\includegraphics[width=5.0in,angle=90,height=3.5in]{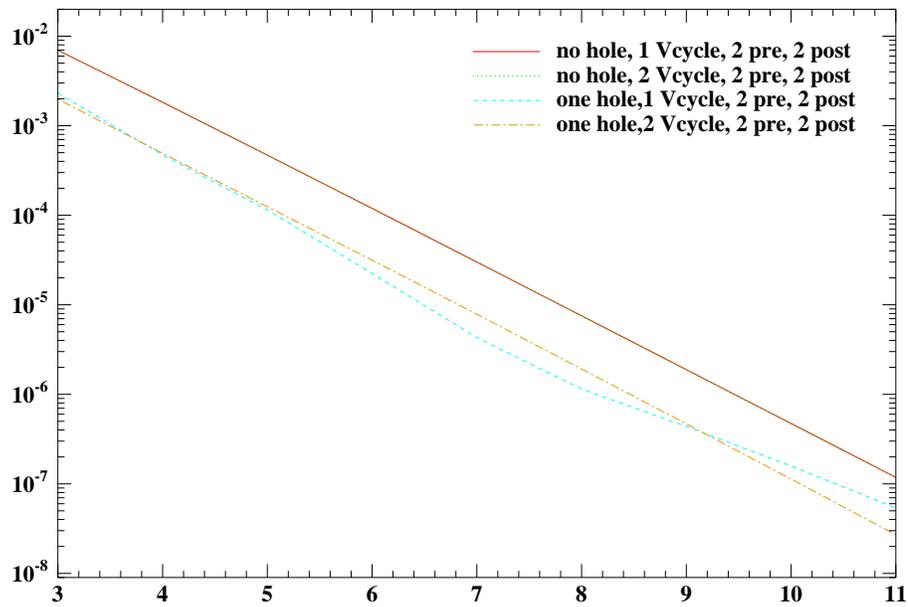}}
\caption{Average error with and without a central hole in the 2D case.
The two ``no hole" lines lie atop one another.\label{fig_err_onehole}}
\end{figure}

Figure \ref{fig_err_onehole} shows the average error over the domain
for 1 and 2 V-cycles (2 pre and 2 post smoothing runs) with and without
a hole.  We need 2 V-cycles to achieve second-order convergence.  As noted
above, the
error for the runs with holes is smaller than in the the base runs (runs
without holes).

\begin{figure}
\centering
\centerline{
\includegraphics[width=5.0in,angle=90,height=3.5in]{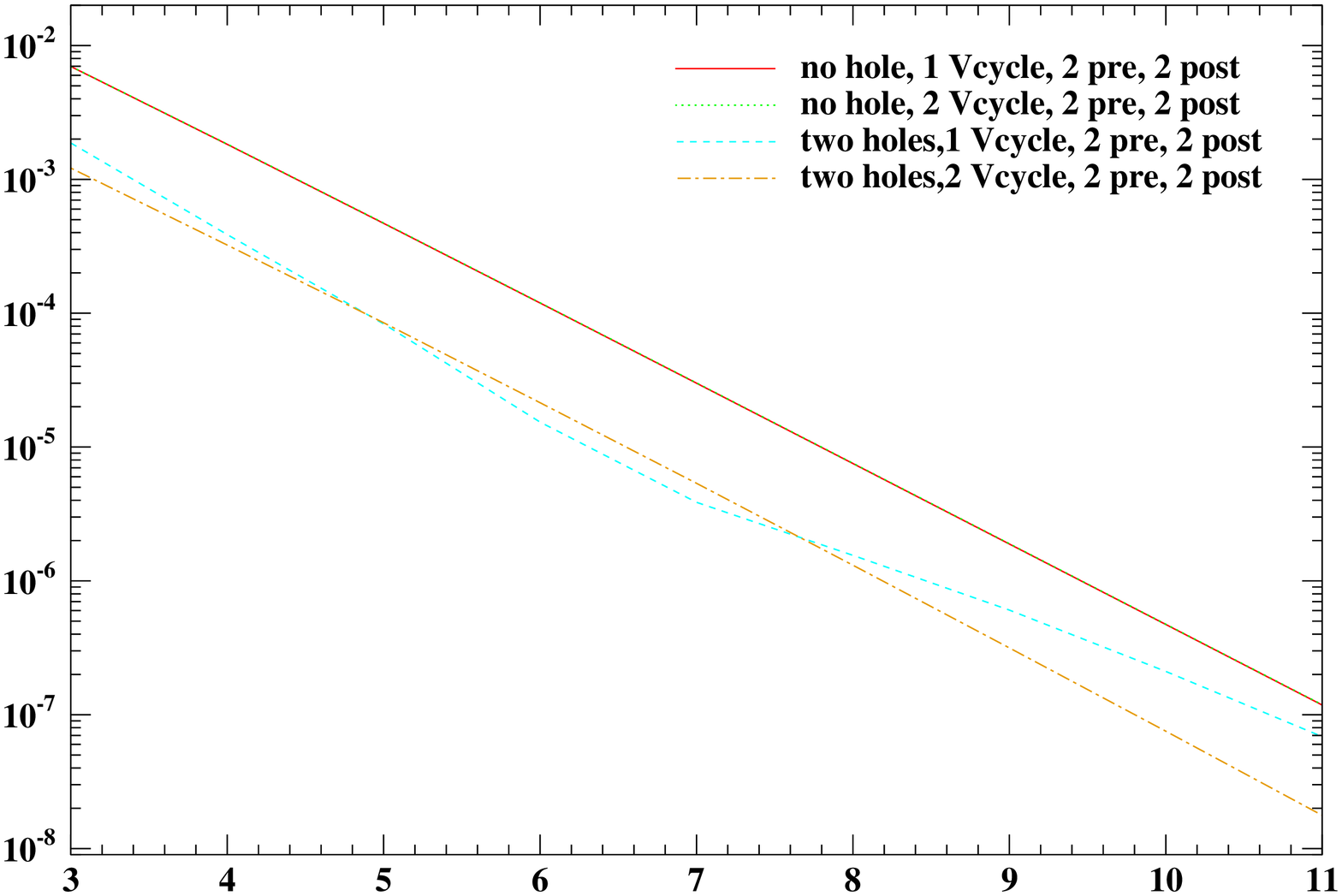}}
\caption{Plot of average error with and without two holes in the 2D case.
The two ``no hole'' lines lie atop one another.\label{fig_err_twohole}}
\end{figure}

Figure \ref{fig_err_twohole} is a similar plot comparing the base runs
with runs over domains with two holes.  Again the error is lower than
the base run without holes.

Figures \ref{fig_2dplot_nohole}
through \ref{fig_err_twohole} demonstrate the dramatic improvement
obtained by our new  inner boundary treatment. In particular
Figure \ref{fig_2dplot_hole} describes the error for precisely the
same problem as that treated in Figure 3 of ref \cite{HawleyMatzner}.
The only difference is the inner boundary treatment. In
(\cite{HawleyMatzner},
Figure 3) the error fluctuates and the maximum occurs near the hole;
maximum error $\sim 0.7 \times 10^{-3}$. The error near the inner boundary
also shows
noticeable imprinting from the cartesian grid. This is achieved for eight
multigrid levels
(finest grid of $2^8+1=257$ vertices along an edge). In
this present work, Figure (\ref{fig_2dplot_hole}), the error near
the hole is quite small and smooth and shows no noticeable rectangular
character.
The maximum error across the grid is $\sim 0.1 \times 10^{-4}$
(more than an order of magnitude smaller than that of
\cite{HawleyMatzner}!). This is accomplished with only {\it seven}
multigrid levels.

Clearly the new inner boundary treatment (inserting new points), with the
important feature of resolving the hole on {\it every} grid,
dramatically outperforms the previous inner boundary method, which
applied boundary conditions only at regular grid points (thus giving a
``rougher'' --
more ``lego-like'' definition of the inner boundary).

While there are many ways this work may be improved,  it is encouraging
that even without mesh refinement it represents a dramatic improvement
in accuracy.

\section{3D Simulations \label{3dsec}}
The Hamiltonian constraint for a single black hole
in a conformally flat background geometry yields the following equation:
\mkeq{
 \nabla^2 u(x,y,z) - K^2 u^5(x,y,z) + {A^2 \over u^7(x,y,z)} = f(x,y,z),
\label{ConstrEq}
}
This is Poisson
equation with two nonlinear terms:
$\nabla^2$ is the usual Laplacian in Euclidean space.  $K^2$
and $A^2$ are arbitrary positive real constants related to the rate
of expansion for the 3-space for which (\ref{ConstrEq}) is the
constraint equation.  $f(x,y,z)$ is related to the energy density
in the 3-space, and can be chosen such that the resulting $u(x,y,z)$
has some known (exact) form by which we can check our numerical
results.  We adjust the matter source term $f(x,y,z)$ such that the
solution is $1 + 2M/r$.  We choose $M=1$ and $K, A = 1$.

For all the 3D simulations we use the domain $[-5.0, -5.0, -5.0]$ to
$[5.0, 5.0, 5.0]$ and a hole of radius $r_{max} = 1.29$.

For the outer boundary in the 3D code to solve the conformally flat 
background, we apply the {\em Robin} condition.  The Robin boundary condition is
$$\frac{\partial (r(u-1))}{\partial r} = 0 $$
This Robin condition is standard condition used in relativity
(\cite{HawleyMatzner}).  We choose to follow Alcubierre
\cite{Miguelcomment} and \cite{HawleyMatzner} and take derivatives
only in direction normal to the faces of our cubical domain.  Since
the stencil for the differencing about a boundary point is not
symmetric about the boundary point we need an extra point to maintain
second order accuracy.  We do so for second order differencing, and for
higher order runs we include the required number points in the stencil
to match the order to that of the interior points.

\hfil\break
\noindent{\em Restriction and Prolongation Operators}\hfil\break

We use the half-weighted restriction operator on normal
interior points, with a weight of $1/12$ for each neighborhood point
along each of the $x,y,z$ direction.
\begin{eqnarray}
           \tilde{u}^{2h}_{I,J,K} & = & \restrict \tilde{u}^h =
              {1\over 2} \tilde{u}^h_{i,j,k} +
              {1\over 12} \left(
                   \tilde{u}^h_{i+1,j,k} + \tilde{u}^h_{i-1,j,k} +
\right.\cr
                   & &\tilde{u}^h_{i,j+1,k} + \tilde{u}^h_{i,j-1,k} +
                   \left. \tilde{u}^h_{i,j,k+1} + \tilde{u}^h_{i,j,k-1}
\right),
\label{3drestrict}
\end{eqnarray}
where $i = 2I - 1$ and $j = 2J -1$. For points near the boundary, 
we adjust the weight to take account of the different spacing.

More complicated restriction operators are often(\cite{HawleyMatzner}) used
which include
weighing all the points on a cube centered around the coarse grid
point.  The results of our simulation are not sensitive to the
restriction operator used, so we prefer the simple half-weighted one
described above.
For the prolongation operator $\prolong$, we continue to use trilinear
interpolation.

\subsection{Second-order convergence}

\begin{figure}
\centering
\centerline{\includegraphics[width=4.0in,angle=90,height=3.5in]{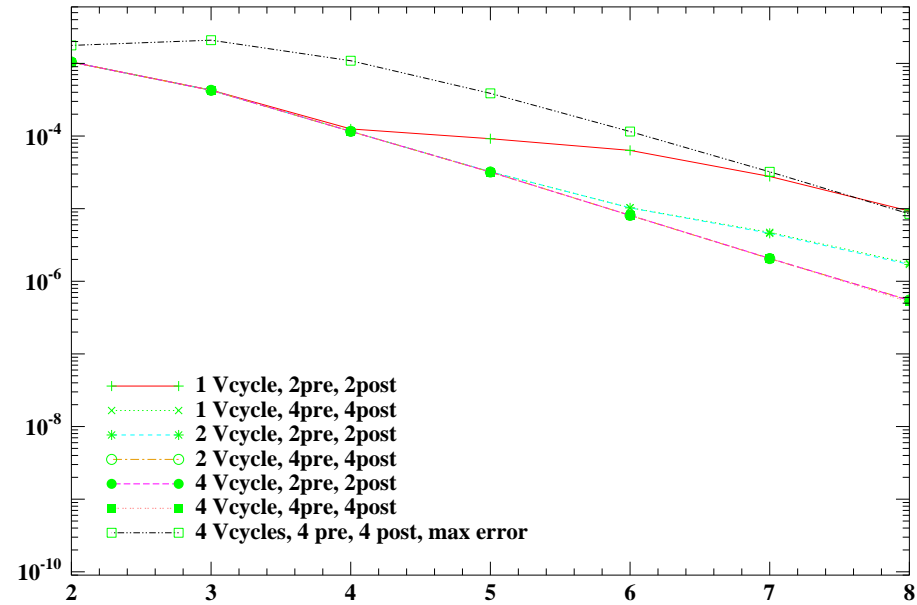}}
\caption{Error norms.  Stencils are chosen for second-order
 convergence and plots with different numbers of V-cycles, pre and post
 smoothings are presented. The two 4 Vcycle average error plots lie atop one another. 
\label{second-order}}
\end{figure}

\hfil\break
\noindent{\em Smoothing Operator}\hfil\break
Again similar to the 2D case, we have to make exceptions for the
interior points near the boundary.  With regular spacing the usual
estimate used to calculate $\nabla^2$ is done as follows; thus
in each direction, (say $x$), the second derivative is
\begin{eqnarray}
 \frac{\partial^2 \tilde{u}}{\partial x^2} & = & h^{-2}
 \left(\tilde{u}_{i+1,j} + \tilde{u}_{i-1,j} - 2 \tilde{u}_{i,j}
 \right) + O(h^2).
\end{eqnarray}
For irregular spacings a simple modification would be
\begin{equation}
 \frac{\partial^2 \tilde{u}}{\partial x^2} =
\frac{2}{h_{i+\frac{1}{2},j}+h_{i-\frac{1}{2},j}}
 \left(\frac{\tilde{u}_{i+1,j} -
   \tilde{u}_{i,j}}{h_{i+\frac{1}{2},j}} + \frac{\tilde{u}_{i-1,j}
   - \tilde{u}_{i,j}}{h_{i-\frac{1}{2},j}}\right) + O(h).
\label{x-dirUneq}
\end{equation}
However because of the irregular spacing in Eq. (\ref{x-dirUneq}) the error
during
smoothing and hence the final solution near the inner boundary will be
first-order
instead of second. (This problem should not exist for
traditional implementations that maintain a regular grid spacing.)
Unlike the 2D case, we have found it essential in the 3D case to have
consistent differencing order at the boundary; otherwise smooth convergence
at
the desired order could not be achieved.
Thus, in order to maintain second-order error bounds near the boundary, we
include an extra interior point.  So a total of four points
($\tilde{u}(x), \tilde{u}(x+h), \tilde{u}(x+2h), \tilde{u}(x+3h) $) are used
to
calculate the second derivative in each of $x,y,z$ coordinates.  Then we
solve the Taylor expansions for $\frac{\partial^2 \tilde{u}}{\partial
 x^2}$.
\begin{eqnarray}
\tilde{u}(x + h_1) & = & \tilde{u}(x)
+ \frac{\partial \tilde{u}}{\partial x}h_1
+ \frac{\partial^2 \tilde{u}}{\partial x^2}\frac{h_1^2}{2}
+ \frac{\partial^3 \tilde{u}}{\partial x^3}\frac{h_1^3}{6}
+ O(h_1^4)\\
\tilde{u}(x + h_2) & = & \tilde{u}(x)
+ \frac{\partial \tilde{u}}{\partial x}h_2
+ \frac{\partial^2 \tilde{u}}{\partial x^2}\frac{h_2^2}{2}
+ \frac{\partial^3 \tilde{u}}{\partial x^3}\frac{h_2^3}{6}
+ O(h_2^4)\\
\tilde{u}(x + h_3) & = & \tilde{u}(x)
+ \frac{\partial \tilde{u}}{\partial x}h_3
+ \frac{\partial^2 \tilde{u}}{\partial x^2}\frac{h_3^2}{2}
+ \frac{\partial^3 \tilde{u}}{\partial x^3}\frac{h_3^3}{6}
+ O(h_3^4)
\end{eqnarray}
As we can see in fig \ref{second-order} we maintain second-order
convergence on all points.

\subsection{Fourth-order convergence}
Given the success in second-order convergence we now
attempt to extend this to fourth order. We find successful
fourth-order convergence by increasing the size of the smoothing stencil to
five points in each ($x,y,z$) direction:
\begin{equation}
 \hspace{-4em}\frac{\partial^2 \tilde{u}}{\partial x^2} = \frac{h^{-2}}{12}
 \left(16(\tilde{u}_{i+1,j} + \tilde{u}_{i-1,j})
   - (\tilde{u}_{i+2,j} + \tilde{u}_{i-2,j}) - 30 \tilde{u}_{i,j}
 \right) + O(h^4).
\end{equation}
Notice that we increase the order of accuracy of the smoothing operator {\it
only}. We do {\it not} modify the interpolation or prolongation operators.

For points near the boundary we need to include one additional point, and
describing
the Taylor expansions.  The Taylor equations for the values over a total of
six
points can then be solved for $\frac{\partial^2
 \tilde{u}}{\partial x^2}$ to $O(h^4)$ accuracy.

The use of stencils that include these additional points could create
problems with the the red-black sweeping of the Gauss smoothing.
Points updated in the red sweep can immediately be used in
calculations for other points within the same sweep. Thus with these
additional points Gauss smoothing can introduce a
bias in the direction of the sweep that could affect the error.
We in fact
observed this bias. We correct this bias as follows, in a sort of
red-black Jacobi method. We assume a red-black labeling of the
vertices (every nearest neighbor of a red is a black, every nearest
neighbor of a black is a red). We then compute the update of each
red vertex, using as many points as appropriate to the the discretization
order, but these updated red values are stored in a separate auxiliary
grid (rather than being written back into the original grid as would
be done in the Gauss procedure). After all red points are updated,
they are all copied back into the original grid. Then the black
points are similarly updated, stored in an auxiliary grid, and
then copied back into the original grid.  The
extra memory required to hold intermediate results is not a concern
since other steps in the multigrid always require extra 
work arrays, which are available for use at this time.

Notice that we increase the order of accuracy of the smoothing operator {\it
only}. We do {\it not} modify the interpolation or prolongation operators. A
further complication is that with larger stencils of high-order methods,
there may not be enough ``room'' on the coarsest grid to construct the
stencil. An obvious solution is to ensure that the coarsest grid {\it is}
sufficiently fine that that the stencils {\it do} fit. Instead, we use the
following approach: If there is insufficient room to construct a stencil at
the
desired order, we drop back to the next highest order that {\it does} fit.
In practice this means the solution on the coarsest grid is always second
(or even first) order.

\begin{figure}
\centering
\centerline{\includegraphics[width=4.0in,angle=90,height=3.5in]{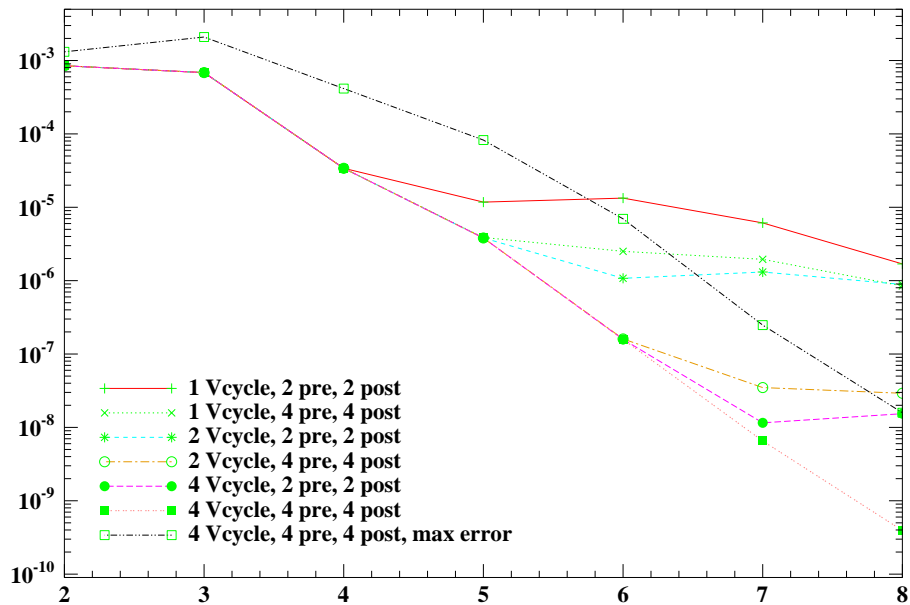}}
\caption{Error norms.  The stencils chosen here should give
 fourth-order convergence.\label{fourth-order}}
\end{figure}

\begin{figure}
\centering
\centerline{\includegraphics[width=4.0in,angle=90,height=3.5in]{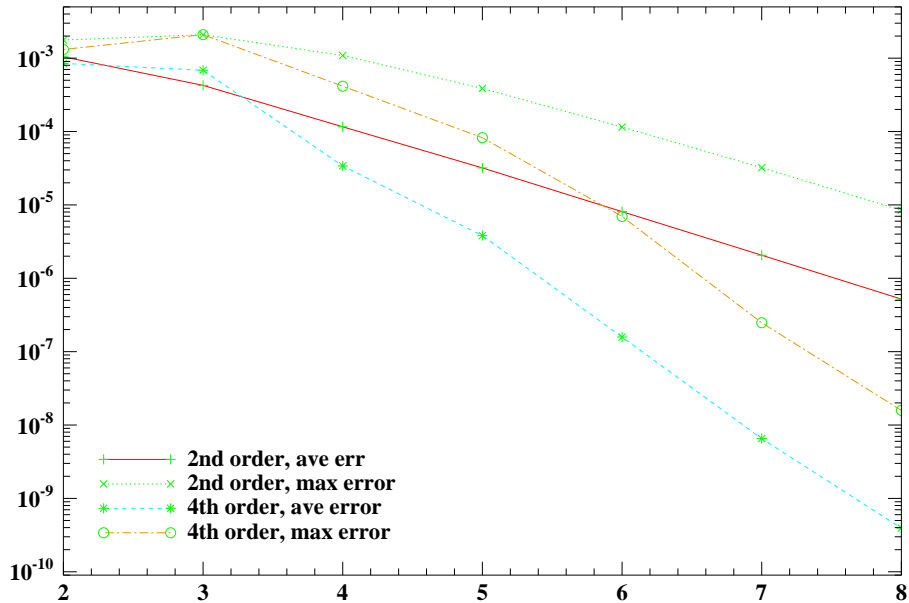}}
\caption{Comparison for second-order and fourth-order, of average error, 
and the maximum error over the
entire domain, which is always located near the inner boundary. The
plot clearly shows that error at the boundary is also controlled to
the correct order\label{comp}}
\end{figure}

Regardless of the simplifications, increasing only the order of the
smoothing, and even dropping back to low-order differencing on
coarse grids, this approach has yielded fourth-order solutions.
Figure \ref{fourth-order} shows the fourth-order convergence we get
with the combination of the new stencils and modified red-black
smoothing.
Figure \ref{comp} shows a comparison between the
second-order and fourth-order methods of the error norms produced.
We do need a larger number of V-cycles and/or smoothing cycles to get
fourth-order
convergence, compared to the second-order case.
We have not yet attempted to reduce or optimize this,
but with fourth-order convergence the error behavior is so much better than
with lower-order
schemes that the extra time spent on additional V-cycles
is not a matter of concern.

\begin{figure}
\centering
\centerline{\includegraphics[width=6.0in,angle=90,height=3.5in]{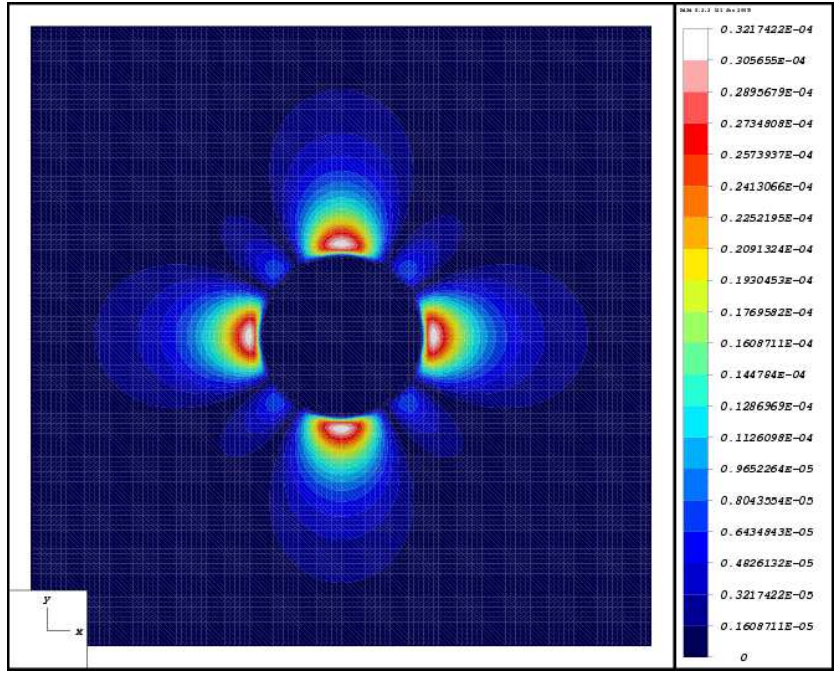}}
\caption{Plot of error over a central slice 
  $x = 0.0$.  Second-order convergence run.
  $l_{\rm max}=7$, $l_{\rm min}=2$, 4 V-cycle and 4 pre- and 4 post-CGC smoothing sweeps
}
\label{xy-3d-second}
\end{figure}

\begin{figure}
\centering
\centerline{\includegraphics[width=6.0in,angle=90,height=3.5in]{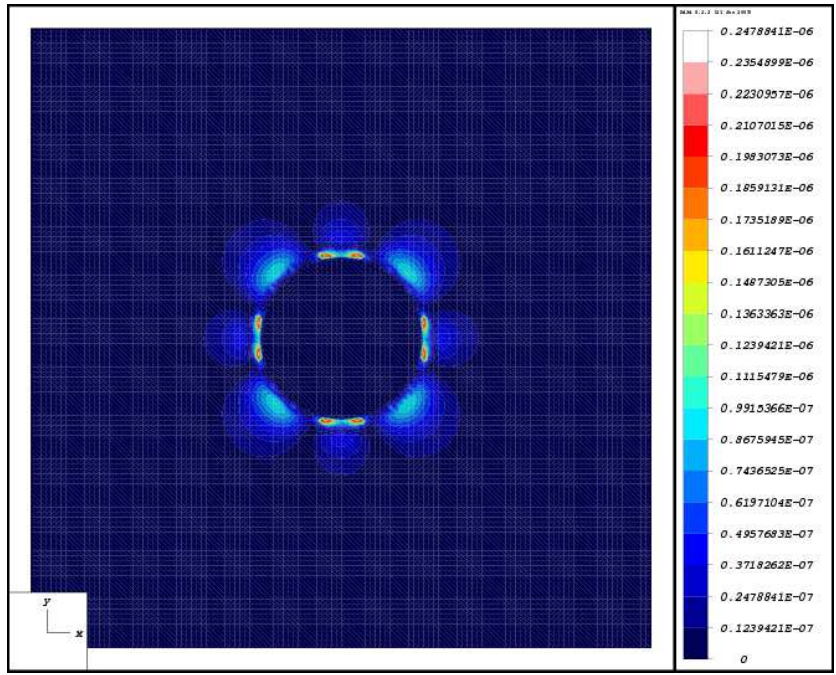}}
\caption{Plot of error over a central slice $x = 0.0$.  Fourth-order run.
$l_{\rm max}=7$, $l_{\rm min}=2$,
4 V-cycle and 4 pre- and 4 post-CGC smoothing sweeps.}
\label{xy-3d-fourth}
\end{figure}

A plot of the errors across the entire domain is useful to visualize
the performance of the second-order and fourth-order methods.  Figures
\ref{xy-3d-second} and \ref{xy-3d-fourth} are plot of the error across the
central slice of the domain.  The error is highest close to the inner
boundary but the fourth-order runs contain the error to a very small
neighborhood of the inner boundary, and the maximum error is a factor $\sim
10^2$ smaller than
that for the second-order solution.

\subsection{Sixth-, and higher-order convergence}
\begin{figure}
\centering\caption{Plot of error norms.\label{high-order}}

\centerline{\includegraphics[width=6.5in,height=5.0in]{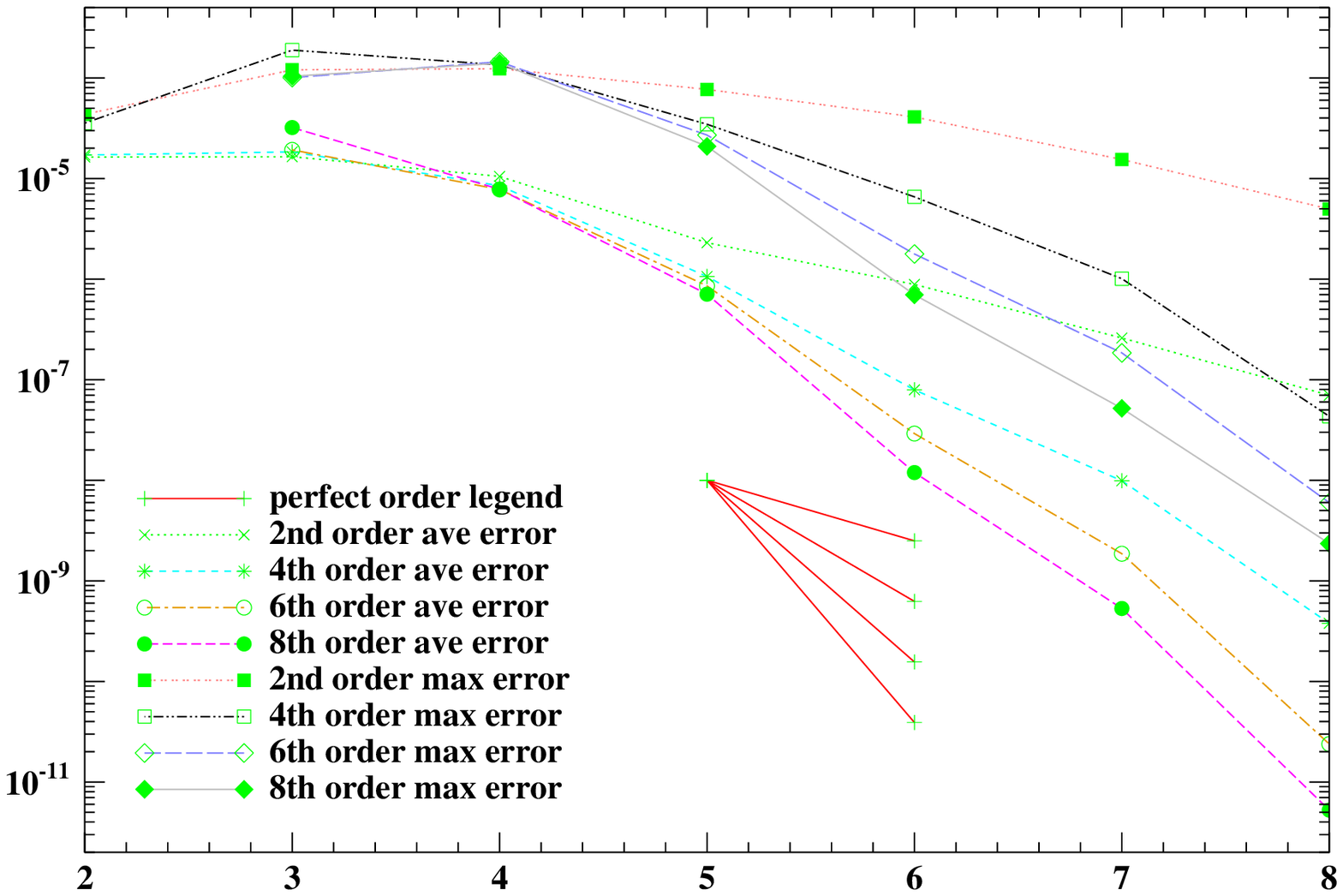}}
\caption{Comparison of average error, and the maximum error over the
entire domain, which is always located near the inner boundary, the
plot clearly shows that error at the boundary is also controlled to
consistent order.}
\label{high}
\end{figure}

The work was further extended to convergence of higher order than fourth.
We include additional points into the stencil to explore sixth- and
eighth-order
convergence.  The results are shown in
Figure \ref{high}. We see that, if a sufficiently large number of levels is
included,
good convergence behavior is obtained to eighth order.  It appears that the number of refinement levels 
to achieve convergence increases with the convergence order. 


\section{Conclusions and further work\label{conc}}
The main ideas presented in this paper are
\begin{itemize}
\item We augment the grid for each level by
 \begin{itemize}
 \item Finding the intersection between every grid line at each level
   with the inner boundaries (holes).
   \item At each of these points we insert a additional boundary grid
     point, whose solution in the Dirichlet boundary conditions is
     known a priori.
 \end{itemize}
We have demonstrated explicitly in the second-order 2D implementation that
this procedure
leads to much improved error at the inner boundary. The essential feature
appears to be resolving the
hole on every grid.
 \item Since the grid spacings near the inner boundaries are now not
   fixed, we modified the  Gauss-Seidel Newton iteration (used to smooth
the error)
   to take one additional neighboring grid point in each of $x,y,z$
   coordinates to maintain the relaxation accuracy.
 \item We achieve good convergence for a second-order implementation. And,
by increasing the
   order of the smoothing operator (only) and even with a drop back to
lower order on the
   coarsest grids, we find good convergence at fourth- and sixth-order.  We
have demonstrated a remarkably simple implementation to provide fourth- and
sixth-order elliptic solutions.
\end{itemize}

Problems with this approach are
\begin{itemize}
\item One of the planned applications of the code is in constrained
evolution.
  The manner in which we introduce grid points leads to some
 grid spacings near the boundary that are much smaller than the regular
 grid spacing. This may cause violations of the
 Courant-Friedrichs-Levy(CFL) constraint on the time steps for
 evolution.
\item The complexity of the data structures and of the coding increases for
higher order
 so that parallelizing this code without introducing bugs might be
difficult.
\end{itemize}

Some of the results of this work have not yet been investigated to
completely understand the behavior.  For example, we do not fully
understand why we need 4 V-cycles for convergence in the 3D runs,
nor have we fully understood the number of levels required to insure higher-order convergence.

\section{Acknowledgments}

This work was supported by NSF grant
PHY~0354842, and by NASA grant NNG04GL37G.

\end{document}